\documentclass[floatfix,aps,preprintnumbers,amsmath,amssymb,nofootinbib,superscriptaddress]{revtex4}
\usepackage[T1]{fontenc}
\usepackage{lmodern}
\usepackage{graphicx}
\usepackage{amsmath,amssymb,bm}
\usepackage{physics}
\usepackage{xcolor}
\usepackage{comment}
\usepackage{multirow}
\usepackage{hyperref}
\usepackage{booktabs}
\hypersetup{colorlinks=true, linkcolor=blue, citecolor=blue, urlcolor=blue}

\newcommand{\bea}{\begin{eqnarray}}
\newcommand{\eea}{\end{eqnarray}}
\newcommand{\be}{\begin{equation}}
\newcommand{\ee}{\end{equation}}

\begin{document}

\title{Scaling properties of two-particle--two-hole responses in
  asymmetric nuclei for neutrino scattering within the
  Relativistic Mean-Field framework}

\author{V.L. Martinez-Consentino}\email[]{victormc@ugr.es}
\affiliation{Departmento de Ciencias Integradas, Universidad de Huelva, E-21071 Huelva, Spain.}
\affiliation{Departamento de Sistemas F\'isicos, Qu\'imicos y Naturales, Universidad Pablo de Olavide, E-41013 Sevilla, Spain}

\author{J.E. Amaro}\email[]{amaro@ugr.es} 
\affiliation{Departamento de F\'{\i}sica At\'omica, Molecular y Nuclear, Universidad de Granada, E-18071 Granada, Spain.}
\affiliation{Instituto Carlos I de F{\'\i}sica Te\'orica y Computacional, Universidad de Granada, E-18071 Granada, Spain.}

\author{J. Segovia}
\email[]{jsegovia@upo.es}
\affiliation{Departamento de Sistemas F\'isicos, Qu\'imicos y Naturales, Universidad Pablo de Olavide, E-41013 Sevilla, Spain}

\date{\today}

\begin{abstract}
We perform a systematic analysis of the nuclear dependence
  of two-particle--two-hole meson-exchange current contributions to
  inclusive lepton-nucleus scattering within the relativistic
  mean-field framework. We present microscopic calculations of nuclear
  responses for a set of 17 nuclei, ranging from helium to uranium,
  using a model with different Fermi momenta for protons and
  neutrons. We propose a novel scaling prescription based on the
  two-particle phase space and key nuclear parameters. The resulting
  description is accurate over a wide range of nuclear targets, with
  typical deviations below 10\%, and allows for a separate treatment
  of the different emission channels. In addition, a consistent
  benchmark against electron-scattering data is provided. The
  parametrization presented provides a practical framework for
  extending the responses to different nuclear targets in neutrino
  event generators.
\end{abstract}

\pacs{25.30.Pt, 25.40.Kv, 24.10.Jv}
\keywords{neutrino scattering, electron scattering, meson-exchange currents, scaling}
\maketitle

\section{Introduction}
The new era of neutrino experiments is characterized by an
unprecedented technological leap. On the one hand, the sophisticated
liquid-argon time projection chamber (LArTPC) technology dominates
promising projects such as DUNE \cite{Dune20a,Dune20b} and the
Short-Baseline Neutrino (SBN) program, which includes experiments such
as MicroBooNE, SBND, and ICARUS \cite{Micro25,SBND20,Icarus23},
establishing $^{40}$Ar as a key nuclear target for present and future
measurements. On the other hand, this experimental landscape is
complemented by large-scale detectors such as Hyper-Kamiokande
\cite{HK}, based on water, and JUNO \cite{JUNO}, whose goal is to
improve the precision of neutrino measurements using hydrocarbon
targets.

Although the field has historically relied on carbon or oxygen as
theoretical reference nuclei for nuclear modeling, the current
experimental program, driven by argon-based detectors and the
multi-target strategy of MINER$\nu$A \cite{Minerva14}, requires going
beyond this simplified framework. As neutrino interactions at the GeV
scale are explored in neutron-rich media, the assumption that heavy
nuclei can be described as simple scaled versions of symmetric nuclei
\cite{Mos16b,Dol19,Sch16} becomes increasingly questionable. This
limitation is particularly evident in the so-called ``dip region'',
located between the quasielastic (QE) peak and the $\Delta(1232)$
resonance.\\

In this kinematic regime, the cross section is dominated by
two-particle--two-hole excitations induced by meson-exchange currents
(2p2h-MEC). A variety of theoretical approaches have been developed to
describe these mechanisms with increasing sophistication
\cite{Nie11,Mos14,Meg16,Van17,Roc18,Mar21a,Rus25}, reflecting their
crucial role in the reliable reconstruction of the incoming neutrino
energy. However, as demonstrated in a recent study focused on nuclei
with $A = 40$ \cite{Mar25b}, neglecting nuclear isospin asymmetry, as
well as the use of same Fermi momenta for protons and neutrons, leads
to non-negligible deviations in the predicted nuclear response. These
discrepancies increase further with increasing mass number.\\

The present work constitutes a natural continuation of that study
\cite{Mar25b} and addresses this issue through a systematic analysis
of 2p2h responses across a broad set of nuclear targets. To this end,
we extend microscopic calculations based on an asymmetric nuclear
matter description within the Relativistic Mean Field (RMF)
framework, previously applied to nuclei such as argon, calcium, and
carbon \cite{Mar23b,Mar25b,Bar19}, to an extensive set of seventeen
nuclei. This set spans from light systems, such as lithium and helium,
to heavy nuclei such as lead and uranium, thereby covering the full
mass range relevant for neutrino scattering experiments, as well as
for inclusive electron scattering for which experimental data exist
for a wide variety of nuclei \cite{archive,archive2}.\\

The main objective of this study is to investigate scaling factors
that allow the prediction of the 2p2h response of an arbitrary nucleus
starting from that of $^{12}$C. In addition, we aim to characterize
the nuclear dependence through the available two-particle phase space
and nuclear parameters such as the number of nucleons, Fermi momenta,
and the effective nuclear mass. The proposed formulation is grounded
in a well-defined theoretical framework based on the factorization of
the nuclear response and explicitly distinguishes both among different
isotopes and among the various emission channels ($pp$, $np$, and
$nn$). This strategy allows for a more controlled extrapolation from
symmetric nuclei to neutron-rich systems, thereby mitigating the
dependence on specific nuclear-model assumptions. To achieve this
goal, the transverse response is described using a common functional
form, with channel-dependent coefficients and phase-space factors, at
fixed momentum transfer.\\

The paper is organized as follows. Section~\ref{formalismo} summarizes
the formalism for the treatment of asymmetric 2p2h-MEC responses
within the RMF framework. In Section~\ref{scaling}, the nuclear
responses are scaled to $^{12}$C and the corresponding proportionality
factors are extracted. Section~\ref{globalfit} presents the factorized
fitting strategy and the residual analysis. Finally,
Section~\ref{conclusiones} summarizes the main conclusions of this
study and outlines future perspectives.

\section{2p2h-MEC Formalism}
\label{formalismo}

\begin{figure}[h!]
\centering
\includegraphics[scale=1.0,bb=-40 10 430 180]{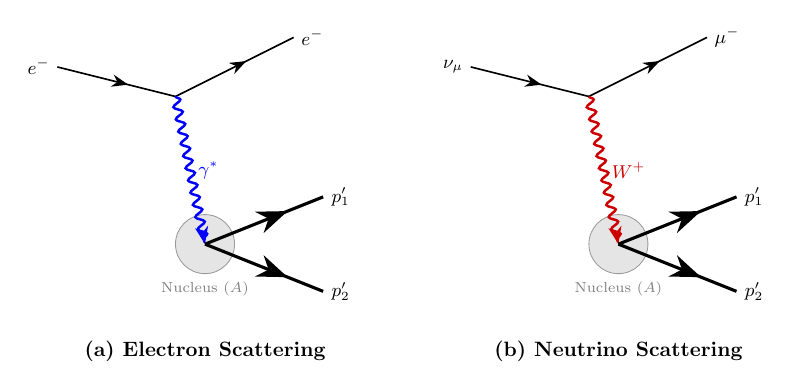}
\caption{Schematic representation of two-particle emission
in lepton-nucleus scattering: electron scattering mediated by the
exchange of a virtual photon (left) and neutrino scattering mediated
by the exchange of a charged $W^{+}$ boson (right).}
\label{diagram}
\end{figure}

The inclusive lepton-nucleus scattering cross section is obtained
from the contraction of the leptonic tensor with the nuclear hadronic
tensor $W^{\mu\nu}$, which encodes the nuclear dynamics. In addition
to one-body processes, further contributions arise from 2p2h
excitations induced by two-body electroweak currents. In these
mechanisms, the exchanged boson interacts with a pair of nucleons,
leading to the emission of two nucleons from the nucleus, as
schematically illustrated in Fig.~\ref{diagram}.

In this work, we follow the formalism introduced in Ref.~\cite{Mar25b}
for asymmetric nuclei, where protons and neutrons are treated
independently by assigning different Fermi momenta, $k_{Fp}$ and
$k_{Fn}$. Within this framework, the contribution to the hadronic
tensor associated with a given two-nucleon emission channel,
corresponding to the nucleonic transition $N_1 N_2 \rightarrow
N'_1N'_2$, can be written as

\begin{align}
W^{\mu\nu}_{N'_1 N'_2} = & \frac{V}{(2\pi)^9}
\int d^3h_1\, d^3h_2\, d^3p'_1\, d^3p'_2\,
\frac{(m_N^*)^4}{E_1 E_2 E'_1 E'_2}\,
w^{\mu\nu}_{N'_1N'_2}
(\mathbf{p}'_1,\mathbf{p}'_2,\mathbf{h}_1,\mathbf{h}_2)
\nonumber\\
& \times
\Theta_{N'_1,N_1}(p'_1,h_1)\,
\Theta_{N'_2,N_2}(p'_2,h_2)\,
\delta(E'_1 + E'_2 - E_1 - E_2 - \omega)\,
\delta(\mathbf{p}'_1 + \mathbf{p}'_2 - \mathbf{h}_1 - \mathbf{h}_2 - \mathbf{q}),
\label{hadronic12}
\end{align}
where $\mathbf{h}_i$ ($\mathbf{p}'_i$) and $E_i$ ($E'_i$) denote the
momenta and energies of the initial (final) nucleons,
with effective mass $m_N^*$, and $V$ is the volume. The delta
functions enforce energy and momentum conservation, with $\omega$ and
$\mathbf{q}$ denoting the energy and momentum transfer,
respectively. The step functions $\Theta_{N',N}(p',h)$ restrict the
phase space to occupied hole states below the corresponding Fermi
momentum and to unoccupied particle states above it, for each isospin
species. The elementary two-body tensor, $w^{\mu\nu}_{N'_1N'_2}$, is
constructed from antisymmetrized two-body current matrix elements as

\begin{equation}
w^{\mu\nu}_{N'_1N'_2}
=
G_s
\sum_{s_1,s_2,s'_1,s'_2}
j^\mu_A(1',2',1,2)^{\ast}\,
j^\nu_A(1',2',1,2),
\end{equation}
where the sum runs over the spin projections of the initial ($s_i$)
and final ($s_i'$) nucleons. The two-body currents $j^\mu_A$ include
pion-exchange mechanisms (seagull and pion-in-flight), as well as
$\Delta$ excitation($\Delta$-forward and $\Delta$-backward). Their
explicit expressions are given in Ref.~\cite{Mar23b}. The factor $G_s$
accounts for the symmetry of the initial/final state, and prevents double
counting in channels involving identical nucleons. For inclusive
neutrino-induced reactions, the factor takes the value $G_s =
1/2$. For electron-induced reactions, $G_s = 1/4$ in the symmetric
$pp$ and $nn$ channels, while $G_s = 1$ in the asymmetric $np$
channel.\\

Medium effects are incorporated through an effective description
inspired by relativistic mean-field models of nuclear matter
\cite{Ser86}.  Within this approach, nucleons propagate with modified
single-particle properties that effectively encode the action of
scalar and vector mean fields in the nuclear medium. The medium
modification is introduced through effective masses for the nucleon
and the $\Delta(1232)$ resonance:

\begin{equation}
m_N^* = m_N - g_s \phi_0,
\qquad
m_\Delta^* = m_\Delta - g_s^\Delta \phi_0 .
\end{equation}
Here, $g_s$ and $g_s^\Delta$ are effective scalar couplings and
$\phi_0$ denotes the mean-field expectation value of the scalar field.
For observables that depend only on energy differences between initial
and final nucleon states, such as the quasielastic 1p-1h response and
the seagull and pion-in-flight 2p2h mechanisms, the contribution of
the vector mean field cancels explicitly. In contrast, mechanisms
involving an intermediate $\Delta$ excitation depend on absolute
energies, so that the vector mean field shifts the $\Delta$ pole and
modifies both the strength and the shape of the corresponding
response.\\

For simplicity, a universal coupling is assumed \cite{Weh93},
assigning the same  vectorial energy to both nucleons and the $\Delta$
resonance, $E_v^N = E_v^\Delta$. The effective nucleon and $\Delta$
masses, together with the proton and neutron Fermi momenta employed
for each nucleus, are summarized in Table~\ref{parametros}. These parameters ($k_{Fp}$, $k_{Fn}$, $m_N^*$, and $m_\Delta^*$) are determined through a phenomenological fit to more than 20000 data points of inclusive electron-nucleus scattering available in the Quasielastic Electron Nucleus Scattering Archive of the University of Virginia \cite{archive, archive2}. The resulting values have been further validated against recent high-precision measurements from JLab \cite{Mur19} and MAMI \cite{Mih24}. The parameters listed in Table \ref{parametros} are consistent with those obtained in previous Superscaling analyses \cite{Mar17, Mar21a}, which employed a single effective Fermi momentum. In this work, we use independent Fermi momenta for protons ($k_{Fp}$) and neutrons ($k_{Fn}$) to account for nuclear asymmetry which is equivalent to the value taken in the previous references. In our procedure, the Fermi momentum and the nucleon effective mass are varied to best describe the quasielastic peak in kinematics where the peak is clearly distinguishable (excluding regions of very low or high momentum transfer). Subsequently, the effective $\Delta$ mass $m_\Delta^*$ is adjusted to reproduce the $\Delta$-resonance maximum, ensuring the peak position remains consistent with Relativistic Fermi Gas predictions \cite{Mar21a}.
\begin{table*}[t]
\caption{Nuclear parameters employed: proton and neutron numbers,
  effective nucleon and $\Delta$ masses normalized, and proton and
  neutron Fermi momenta (in MeV/$c$).}
\label{parametros}
\begin{tabular*}{\textwidth}{l@{\extracolsep{\fill}}cccccc}
\toprule
Nucleus & $Z$ & $N$ & $m_N^*/m_N$ & $m_\Delta^*/m_\Delta$ & $k_{Fp}$ & $k_{Fn}$ \\
\midrule
$^4$He    & 2  & 2  & 0.90 & 0.922 & 160 & 160 \\
$^6$Li    & 3  & 3  & 0.80 & 0.846 & 165 & 165 \\
$^9$Be    & 4  & 5  & 0.80 & 0.846 & 190 & 205 \\
$^{12}$C  & 6  & 6  & 0.80 & 0.846 & 225 & 225 \\
$^{16}$O  & 8  & 8  & 0.80 & 0.846 & 230 & 230 \\
$^{24}$Mg & 12 & 12 & 0.75 & 0.807 & 235 & 235 \\
$^{27}$Al & 13 & 14 & 0.80 & 0.846 & 235 & 240 \\
$^{40}$Ar & 18 & 22 & 0.73 & 0.792 & 237 & 256 \\
$^{40}$Ca & 20 & 20 & 0.73 & 0.792 & 240 & 240 \\
$^{48}$Ca & 20 & 28 & 0.73 & 0.792 & 240 & 268 \\
$^{56}$Fe & 26 & 30 & 0.70 & 0.770 & 236 & 248 \\
$^{59}$Ni & 28 & 31 & 0.67 & 0.747 & 234 & 242 \\
$^{89}$Y  & 39 & 50 & 0.65 & 0.732 & 232 & 252 \\
$^{119}$Sn& 50 & 69 & 0.65 & 0.732 & 230 & 256 \\
$^{181}$Ta& 73 & 108 & 0.65 & 0.732 & 228 & 260 \\
$^{208}$Pb& 82 & 126 & 0.65 & 0.732 & 230 & 265 \\
$^{238}$U & 92 & 146 & 0.65 & 0.732 & 252 & 293 \\
\bottomrule
\end{tabular*}
\end{table*}

Following the integration procedure described in Ref.~\cite{Sim17},
momentum conservation is used to perform the integration over
$\mathbf{p}'_2$, and rotational invariance around the direction of the
momentum transfer is exploited. The resulting nuclear response
functions can be expressed as

\begin{align}
R^{K}_{N'_1 N'_2}
=
\frac{V}{(2\pi)^9}
\int
d^3p'_1\, d^3h_1\, d^3h_2\,
\frac{(m_N^*)^4}{E_1 E_2 E'_1 E'_2}\,
r^{K}(\mathbf{p}'_1,\mathbf{p}'_2,\mathbf{h}_1,\mathbf{h}_2)
\nonumber\\
\times
\Theta_{N'_1,N_1}(p'_1,h_1)\,
\Theta_{N'_2,N_2}(p'_2,h_2)\,
\delta(E'_1 + E'_2 - E_1 - E_2 - \omega),
\label{responses}
\end{align}
where $K$ labels the different response channels and $r^{K}$ denotes
the corresponding reduced response function. The integration over the
energy-conserving Dirac delta is performed in the center-of-mass
system. For electromagnetic interactions, only the longitudinal and
transverse responses contribute:

\begin{equation}
R^L_{\mathrm{em}} = W^{00}_{\mathrm{em}}, \qquad
R^T_{\mathrm{em}} = W^{11}_{\mathrm{em}} + W^{22}_{\mathrm{em}},
\label{relem}
\end{equation}
with allowed final-state channels $pp$, $np$, and $nn$. For
charged-current neutrino reactions, five independent response
functions contribute:

\begin{align}
R^{CC} &= W^{00}, \quad
R^{CL} = -\tfrac{1}{2}(W^{03}+W^{30}), \quad
R^{LL} = W^{33}, \nonumber\\
R^{T}  &= W^{11}+W^{22}, \quad
R^{T'} = -\tfrac{i}{2}(W^{12}-W^{21}),
\label{rneu}
\end{align}

\section{Scaling of 2p2h Responses Relative to Carbon}
\label{scaling}
We compute the 2p2h nuclear response functions for a set of seventeen
nuclei at a fixed momentum transfer $q = 500$~MeV/$c$, corresponding
to a kinematic region where the nuclear response is close to its
maximum and which is relevant for modern neutrino experiments. For
each nuclear target $X$ and response channel $K$, we define the 2p2h
scaling ratio relative to ${}^{12}$C as
\begin{equation}
  \mathcal{R}^{K}_{N'_1 N'_2}(X;\omega)\equiv 
\frac{R^{K}_{N'_1 N'_2}(X;q=500,\omega)}
     {R^{K}_{N'_1 N'_2}({}^{12}\mathrm{C};q=500,\omega)}
\approx  \text{const.}
\label{scaling_ratio}
\end{equation}
As discussed in Ref.~\cite{Mar25b}, this ratio exhibits only a weak
dependence on both the response channel $K$ and the momentum transfer.
In that reference, it is shown that the ratio evaluated at
$q=1000$~MeV/$c$ is very similar to that obtained at
$q=500$~MeV/$c$. This behavior is also
consistent with other parametrizations available in the literature
\cite{Sch16,Mos16b,Dol19}, where the scaling is found to be largely
independent of both the momentum transfer and the specific response
considered. Accordingly, in the present analysis we restrict ourselves
to the transverse channel and drop the index $K$, so that the ratio is
simply denoted as $\mathcal{R}$.\\

In practice, for each nucleus we extract a single effective scaling
factor by evaluating the transverse responses $R^T$ at their
respective maxima and taking the ratio of peak values, dividing the
maximum of the nucleus $X$ by that of ${}^{12}\mathrm{C}$. The peak
positions occur at very similar energy transfers, ensuring a
well-defined normalization. The scaling ratios are therefore
evaluated by taking $\mathcal{R}$ at the maximum of the transverse
response, which for $q=500$~MeV/$c$ lies around $\omega \simeq
300$-$400$~MeV.\\

\begin{figure}[h]
\centering
\includegraphics[scale=0.95]{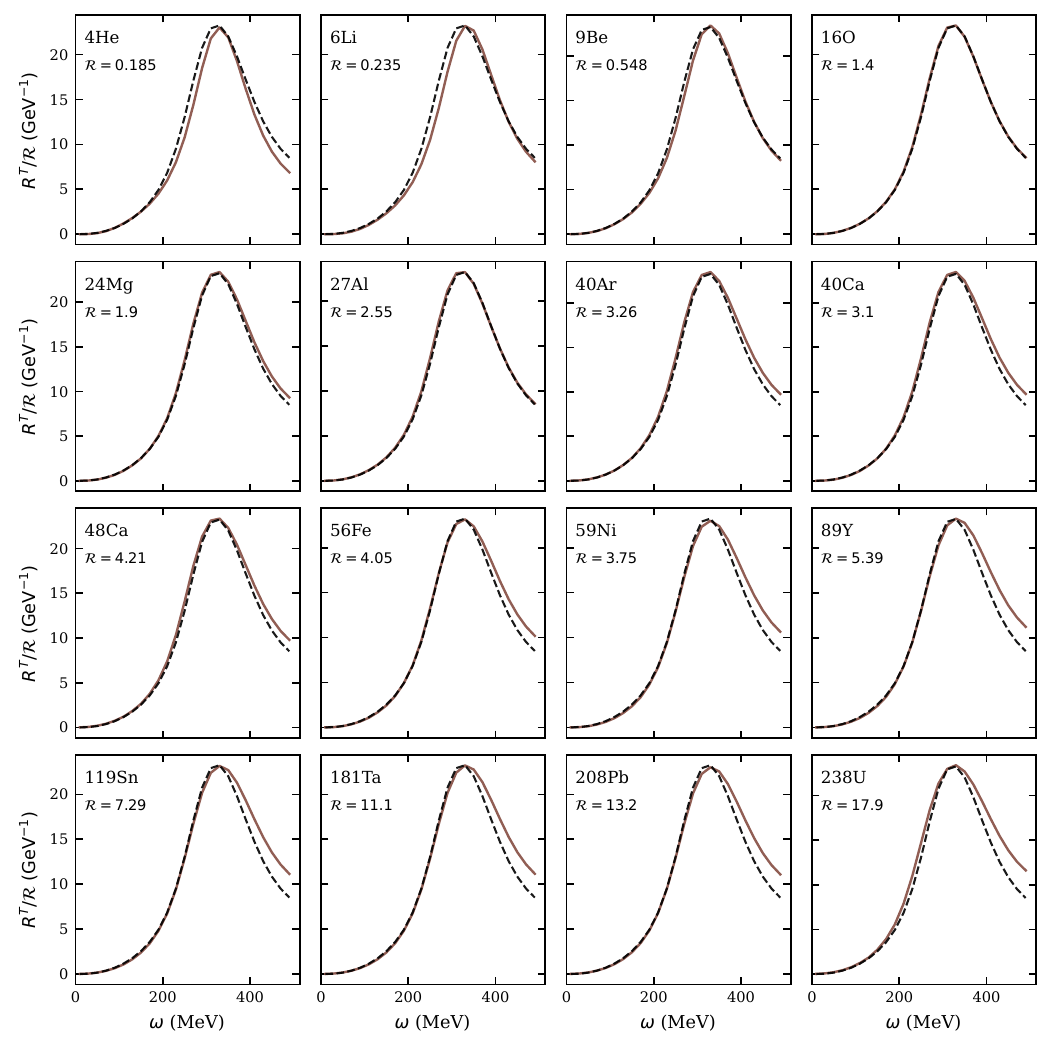}
\caption{Transverse 2p2h responses for neutrino-induced
$np \rightarrow pp$ emission at $q = 500$~MeV/$c$.
Solid colored lines correspond to each nucleus, while the dashed black
line shows the ${}^{12}$C result. Each panel displays the response
rescaled by the factor $\mathcal{R}$ indicated in the
figure.}
\label{fig_nu_pp}
\end{figure}

\begin{figure}[h]
\centering
\includegraphics[scale=0.95]{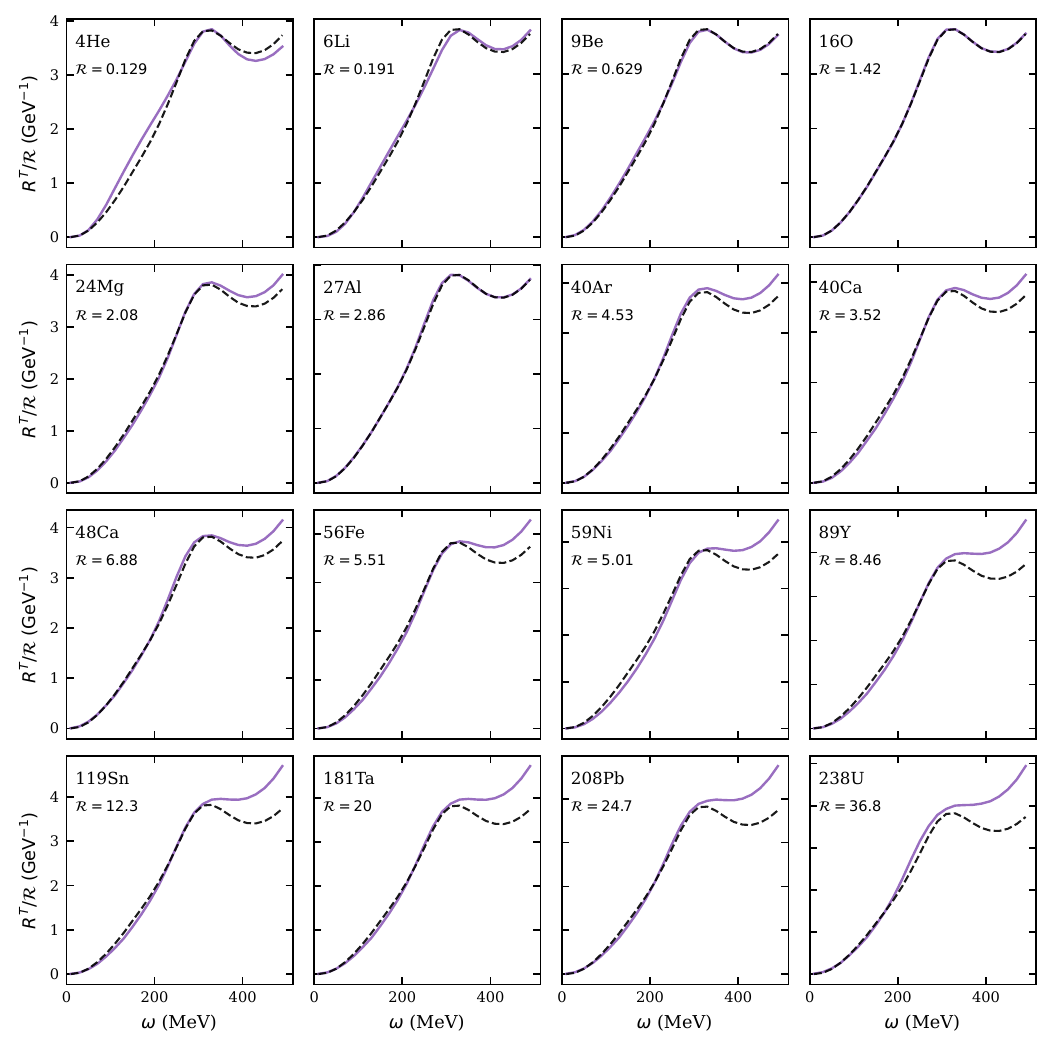}
\caption{Same as Fig.~\ref{fig_nu_pp}, but for neutrino-induced
$nn \rightarrow np$ emission.}
\label{fig_nu_np}
\end{figure}

Figures~\ref{fig_nu_pp} and \ref{fig_nu_np} display the transverse 2p2h
responses for neutrino-induced reactions in the $np \to pp$ and
$nn \to np$ emission channels, respectively, for the seventeen nuclei
considered. The $^{12}$C result is shown as a dashed black curve,
whereas the other nuclei are rescaled by dividing by their corresponding
scaling factor $\mathcal{R}$, quoted in each subpanel.

All responses exhibit a pronounced peak as a function of $\omega$,
consistent with the leading role of $\Delta$-driven MEC dynamics in
this kinematic domain. In the $np \to pp$ channel
(Fig.~\ref{fig_nu_pp}), the rescaled curves show a good collapse around
the carbon reference. The lightest nuclei ($^{4}$He, $^{6}$Li,
$^{9}$Be) display small residual differences, including a mild shift
of the peak region with respect to $^{12}$C. Within the RMF-inspired
effective description, variations of the effective nucleon and
$\Delta$ masses and of the corresponding Fermi momenta
induce small changes in the peak position and width, effects that are
more visible in light systems.\\

Intermediate-mass nuclei provide the most stable behavior: oxygen,
magnesium, and aluminum lie very close to the carbon reference over the
peak region. This point matters for applications, since $^{40}$Ar and
$^{40}$Ca also follow the same pattern around the maximum, supporting
controlled extrapolations from carbon to argon-based detectors.

For heavy nuclei (tin, lead, uranium), the spread increases, especially
at energy transfers above the $\Delta$-peak maximum. Even so, the
description remains adequate when one accounts for the strong growth of
the absolute 2p2h strength with mass number.\\

In the $nn \to np$ channel (Fig.~\ref{fig_nu_np}), we find the same
overall trend, but with an absolute magnitude roughly six times smaller
than in the $np \to pp$ channel. Here the scaling is visibly less
accurate, consistent with the enhanced sensitivity of the $nn \to np$
response to neutron excess. The peak region is
also broader and less sharply defined. This behavior follows from the
interplay of $\Delta$ mechanisms: the $\Delta$-forward diagram
dominates channels with two protons in the final state, whereas the
$\Delta$-backward contribution becomes comparatively more relevant in
the $np$ final channel, producing an effective smearing in the
$\omega$ dependence, in line with the analysis in Ref.~\cite{Mar24b}.
In both channels, $\mathcal{R}$ increases generally with atomic number, tracking
the approximate proportionality of the 2p2h strength to the nuclear
volume. In Appendix~\ref{apendice1}, we compare the corresponding
transverse electromagnetic $T$ responses for electron-induced
reactions in the $nn$, $np$, and $pp$ channels, using coefficients
different from those employed for neutrinos.\\

\begin{figure}[h]
\centering
\includegraphics[scale=0.95]{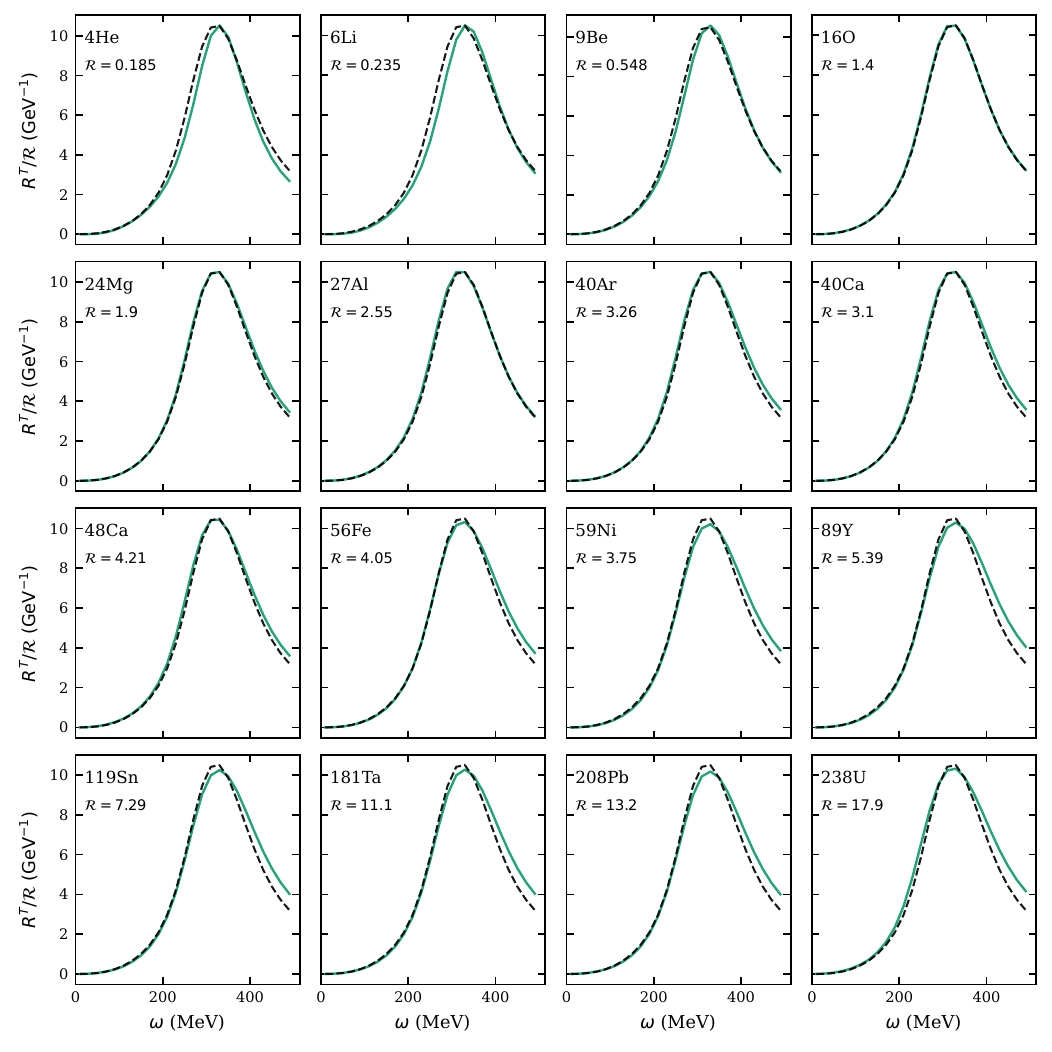}
\caption{Response $R^{T'}$ rescaled to carbon for
neutrino-induced $np \rightarrow pp$ emission at
$q = 500$~MeV/$c$. Solid colored lines correspond to each nucleus, while
the dashed black line shows the ${}^{12}$C result. Each panel displays
the response rescaled by the  factor $\mathcal{R}$ indicated
in the figure.}
\label{fig_nu_pp_tp}
\end{figure}

\begin{figure}[h]
\centering
\includegraphics[scale=0.95]{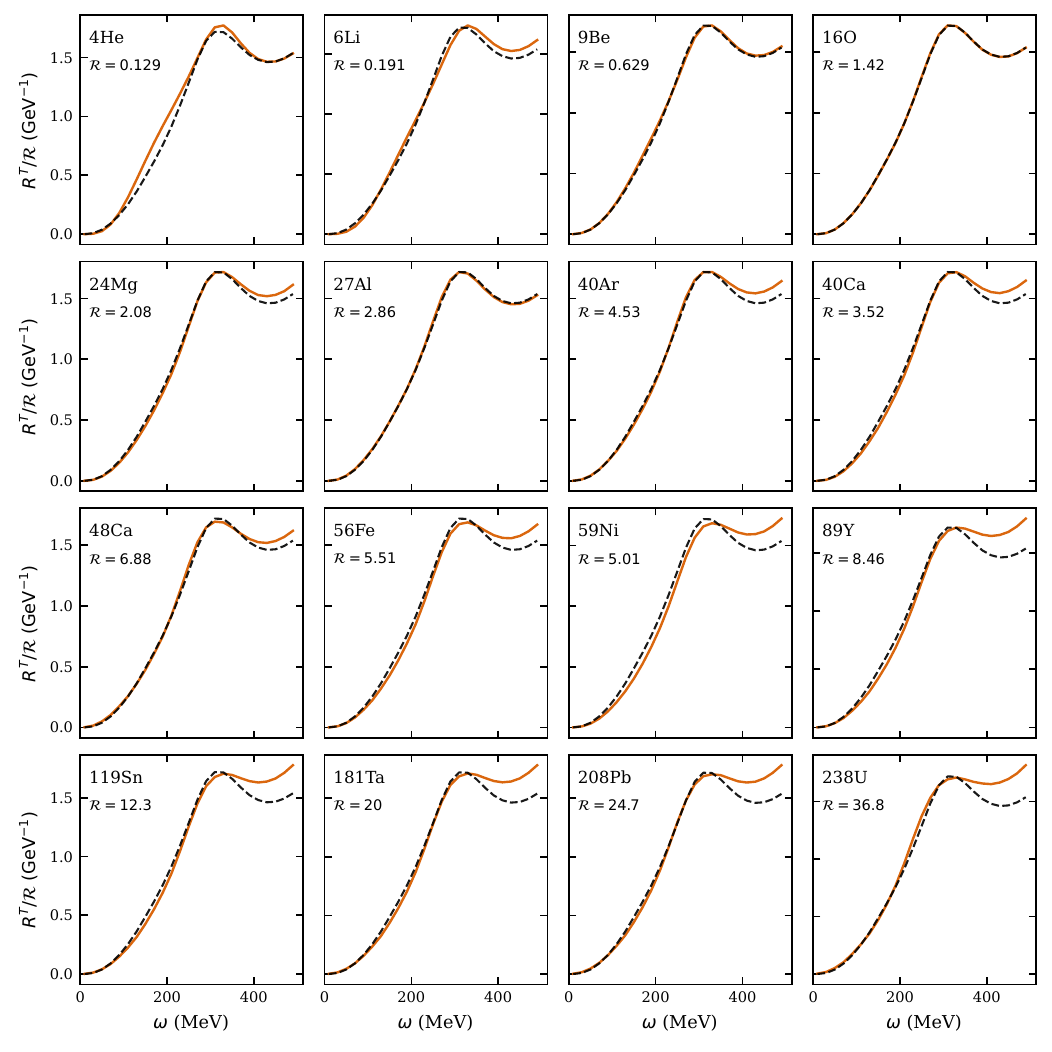}
\caption{Same as Fig.~\ref{fig_nu_pp_tp}, but for neutrino-induced
$nn \rightarrow np$ emission.}
\label{fig_nu_np_tp}
\end{figure}

In Figs.~\ref{fig_nu_pp_tp} and \ref{fig_nu_np_tp} we apply the same
scaling analysis to the $T'$ response in neutrino-induced reactions.
The scaling factor is extracted from the $T$ response and then used
for the remaining neutrino responses ($CC$, $CL$, $LL$, and
$T'$). In the energy range considered, $T$ and $T'$ dominate the
neutrino cross section. We also obtain the approximate factor
$\sim 1/2$ between $T$ and $T'$, as reported in
Ref.~\cite{Mar21b}. Overall, the rescaled $T$ and $T'$ responses
follow the same pattern, with small channel-dependent differences. For
example, in the $pp$ final-state channel for $^{208}$Pb the maximum
ratio extracted from $T'$ is slightly larger than the reference value.\\

Nevertheless, this analysis provides a useful test to assess whether
the scaling behavior identified for the transverse channel also
extends to other response functions. This aspect is particularly
relevant in the antineutrino case, since the $T'$ response changes
sign, thereby reducing the cross section and increasing the relative
contribution of the other responses.\\

Table~\ref{ratios} summarizes the scaling ratios $\mathcal{R}$ for
all nuclei and emission channels, normalized to unity for $^{12}$C
at $q=500$~MeV/$c$. The ratios grow monotonically with nuclear
mass, mirroring the strong increase of the overall 2p2h strength with
nuclear size. For a fixed nucleus, channels involving $nn$ pairs
typically give the largest $\mathcal{R}$, followed by $np$, while
$pp$ channels show the weakest mass dependence. The fact that
neutrino and electron induced reactions exhibit similar global trends
supports the use of $\mathcal{R}$ as a stable, nucleus-dependent
scaling factor.

A striking feature is the much steeper mass dependence of the
$\nu(np)$ and $e(nn)$ channels. From helium to uranium, these
ratios increase by a factor of about 300, whereas the remaining
channels grow by $\sim 100$-$120$. This trend is quantified by the
scaling ratios reported in Table~\ref{ratios}, which span more than an
order of magnitude across the nuclear chart. For example, in the
neutrino-induced $np \to pp$ channel the ratio increases from
$\mathcal{R} \simeq 0.185$ for $^{4}$He to $\mathcal{R} \simeq
18.5$ for $^{238}$U, whereas in the $nn \to np$ channel it
reaches $\mathcal{R} \simeq 38.5$ for $^{238}$U (all values
normalized to unity for $^{12}$C).\\

\begin{table*}[t]
\caption{Scaling ratios $\mathcal{R}$ for different nuclei and emission
channels. The labels $pp$, $np$, and $nn$ refer to the final-state
nucleon pair. All values are normalized to unity for ${}^{12}$C at
$q=500$~MeV/$c$.}
\label{ratios}
\begin{tabular*}{\textwidth}{l@{\extracolsep{\fill}}ccccc}
\toprule
Nucleus & $\nu (pp)$ & $\nu (np)$ & $e (pp)$ & $e (np)$ & $e (nn)$ \\
\midrule
$^4$He     & 0.185  & 0.129  & 0.117  & 0.181  & 0.117 \\
$^6$Li     & 0.235  & 0.191  & 0.160  & 0.234  & 0.160 \\
$^9$Be     & 0.518  & 0.629  & 0.364  & 0.547  & 0.597 \\
$^{12}$C   & 1.000  & 1.000  & 1.000  & 1.000  & 1.000 \\
$^{16}$O   & 1.399  & 1.422  & 1.437  & 1.399  & 1.437 \\
$^{24}$Mg  & 1.901  & 2.075  & 2.062  & 1.931  & 2.062 \\
$^{27}$Al  & 2.547  & 2.859  & 2.507  & 2.537  & 2.928 \\
$^{40}$Ar  & 3.257  & 4.530  & 3.012  & 3.279  & 4.566 \\
$^{40}$Ca  & 3.096  & 3.521  & 3.484  & 3.165  & 3.484 \\
$^{48}$Ca  & 4.210  & 6.881  & 3.484  & 4.172  & 6.976 \\
$^{56}$Fe  & 4.050  & 5.508  & 3.914  & 4.082  & 5.251 \\
$^{59}$Ni  & 3.755  & 5.011  & 3.759  & 3.798  & 4.636 \\
$^{89}$Y   & 5.389  & 8.457  & 4.772  & 5.482  & 7.874 \\
$^{119}$Sn & 7.288  & 12.32  & 5.950  & 7.331  & 11.39 \\
$^{181}$Ta & 11.10  & 20.03  & 8.442  & 11.12  & 18.65 \\
$^{208}$Pb & 13.24  & 24.73  & 9.758  & 13.07  & 23.22 \\
$^{238}$U  & 17.90  & 36.79  & 14.48  & 17.76  & 35.95 \\
\bottomrule
\end{tabular*}
\end{table*}

\begin{figure}[h!]
\centering 
\includegraphics[scale=0.95]{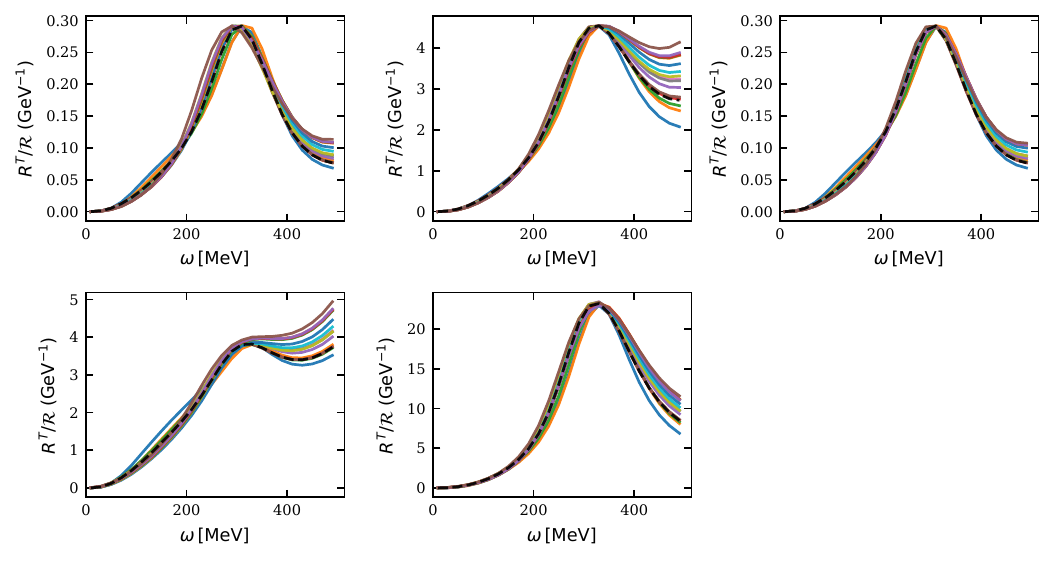}
\caption{Transverse responses $R^T(X)$ rescaled to ${}^{12}$C by dividing
by the corresponding scaling ratio $\mathcal{R}$, for the seventeen
nuclei considered at $q = 500$~MeV/$c$. The dashed black lines represent the results for $^{12}$C, while the different colored lines correspond to the various rescaled nuclei considered in this work. \textbf{Top}: $pp$, $np$, and
$nn$ channels in electron scattering. \textbf{Bottom}: $np$ (left) and $pp$ (right)
final-state channels in neutrino scattering. }
\label{scaling2p2h}
\end{figure}

Figure~\ref{scaling2p2h} collects all transverse responses rescaled to
$^{12}$C for both neutrino and electron reactions. After
applying the inverse factors $1/\mathcal{R}$, the results for all
nuclei fall within a narrow band around the carbon reference, showing
that the leading nuclear dependence can be factorized to a good
approximation. The collapse is tight on the low-$\omega$ side and near
the $\Delta$ peak, where the dominant 2p2h strength is concentrated,
while a larger spread appears at higher energy transfer, where some
nuclei do not scale as well. The $np$ channel shows the largest
residual deviations in both electron- and neutrino-induced cases,
pointing to a stronger channel-specific nuclear dependence.\\

We emphasize that this collapse does not imply exact scaling, but rather
a similar behavior of the rescaled responses across nuclei, with
visible residual nuclear dependence that falls within the typical
uncertainties associated with 2p2h responses.

\section{Factorized Parametrization of the 2p2h Nuclear Response}
\label{globalfit}

In this section we introduce the factorized parametrization used to
capture the nuclear dependence of the 2p2h responses. The input is the
set of scaling ratios listed in Table~\ref{ratios}, extracted from the
microscopic RMF calculations at $q=500~\mathrm{MeV}/c$ and at the
energy transfer corresponding to the maximum of the $\Delta$ peak,
$\omega=\omega_\Delta^{\rm max}$.

Previous studies, such as Ref.~\cite{Lal14,Mar21b,Mar25b}, have shown
that the scaling behavior of the 2p2h response can be partially
interpreted in terms of a phase-space proportionality. While this
approach provides a reasonable description for nuclei with similar
mass numbers (e.g., ${}^{12}$C and ${}^{40}$Ca), significant
deviations emerge for heavier systems such as lead or uranium,
indicating that additional sources of nuclear dependence must be taken
into account. Motivated by these observations, we decompose the
nuclear response in Eq.~\eqref{responses} within a factorized scheme
of the form
\begin{equation}
R^{K}_{N'_1 N'_2}(X;q,\omega)\ =
\
V(X)\,F_{N_1N_2}(X;q,\omega)\,
\frac{\langle r^{K}_{N'_1N'_2}(X;q,\omega)\rangle}{(2\pi)^9}\,,
\label{factorizedeq}
\end{equation}
It is important to emphasize that Eq.~\eqref{factorizedeq} does not introduce
any approximation. It follows exactly from the definition of the averaged
reduced response $\langle r \rangle$. This factorized form isolates a
volume-like contribution $V(X)$, a reduced phase-space factor
$F_{N_1N_2}(X)$, and an averaged reduced single-pair response
$r^{K}_{N'_1N'_2}(X)$, which contains the remaining dynamical and
many-body nuclear dependence. We define the scaling ratio relative
to $^{12}$C as
\begin{equation}
\mathcal{R}(X)\equiv
\frac{R^{K}_{N'_1 N'_2}(X)}{R^{K}_{N'_1 N'_2}({}^{12}\mathrm{C})}
=
\nu(X)\,\varphi_{N_1N_2}(X)\,\rho(X)\,,
\label{R_factorized}
\end{equation}
with
\begin{equation}
\nu(X) \equiv \frac{V(X)}{V({}^{12}\mathrm{C})},\qquad
\varphi_{N_1N_2}(X) \equiv \frac{F_{N_1N_2}(X)}{F_{N_1N_2}({}^{12}\mathrm{C})},\qquad
\rho(X) \equiv \frac{r^{K}_{N'_1N'_2}(X)}{r^{K}_{N'_1N'_2}({}^{12}\mathrm{C})}\,.
\label{definiciones}
\end{equation}
In our implementation, the volume-like factor entering the normalization
is defined under the assumption that protons and neutrons occupy the
same spatial volume within the nucleus. Within this 
picture, the proton and neutron densities differ through their respective
Fermi momenta, but the geometric volume is taken to be common. Under this
assumption, the volume factor can be written equivalently as
\begin{equation}
\frac{V(X)}{(2\pi)^3}
=
\frac{3\,Z}{8\pi\,k_{Fp}^3}
=
\frac{3\,N}{8\pi\,k_{Fn}^3}\,,
\end{equation}
so that $\nu(X)$ is fully determined by $Z$ (or by $N$),
and the corresponding Fermi momentum.

The factor $F_{N_1 N_2}(X)$ represents the 2p2h phase-space contribution
with the volume term removed. Although it is formally defined as a
seven-dimensional integral, it can be accurately approximated by an
analytical expression obtained within the frozen approximation and by
neglecting Pauli blocking effects, as discussed in
Eqs.~(39) and (40) of Ref.~\cite{Sim14}. Within this approximation, the
phase-space factor can be rewritten in the RMF framework allowing for
different proton and neutron Fermi momenta as

\begin{equation}
F(q,\omega)_{N_1 N_2}
=
4\pi
\left( \frac{4}{3}\pi k_{F N_1}^3 \right)
\left( \frac{4}{3}\pi k_{F N_2}^3 \right)
\frac{m_N^{*2}}{2}\,
\frac{p''_1}{E''_1} \,,
\label{analytical}
\end{equation}
where the ratio $p''_1/E''_1$ is given by
\begin{equation}
\frac{p''_1}{E''_1}
=
\sqrt{1-\frac{4m_N^{*2}}{(2m_N^*+\omega)^2-q^2}} \,.
\end{equation}
This expression shows explicitly that the available two-particle phase
space grows rapidly with increasing energy transfer $\omega$ and
approaches a constant value in the asymptotic limit $\omega
\rightarrow \infty$, leading to
\begin{equation}
F(q,\infty)_{N_1 N_2}
=
4\pi
\left( \frac{4}{3}\pi k_{F N_1}^3 \right)
\left( \frac{4}{3}\pi k_{F N_2}^3 \right)
\frac{m_N^{*2}}{2} \, .
\end{equation}
In the present implementation, we factor out only the leading
dependence on the proton and neutron Fermi momenta. Any remaining
kinematic and dynamical dependence either cancels explicitly when
forming the scaling ratio or is retained in the reduced response. One
example of the latter is the effective nucleon mass, which also enters
the phase-space structure and whose contribution is ultimately
absorbed into the reduced-response scaling function
$\rho(X)$. Accordingly, the phase-space term that we factorize for
each emission channel is taken to be

\begin{align}
F_{np} &\propto k_{Fp}^3\,k_{Fn}^3,\qquad
F_{pp} \propto k_{Fp}^6,\qquad
F_{nn} \propto k_{Fn}^6,
\end{align}
While the volume and phase-space factors can be computed using
analytical expressions, the reduced response involves a much more
complex dynamical structure. For this reason, all the remaining
nuclear dependence is collected in the reduced-response scaling
function term $\rho(X)$. We model $\rho(X)$ with an ansatz that
encodes (i) proton--neutron imbalance through the Fermi momenta
$k_{Fp}$ and $k_{Fn}$, and (ii) RMF medium effects through the nucleon
effective mass. Note that, once the leading volume dependence has been
factored out, the only quantities that vary from nucleus to nucleus in
this model are the proton and neutron Fermi momenta and the effective
mass.  Accordingly, we parametrize $\rho(X)$ as

\begin{equation}
\rho(X)=
1
+b_1\left(\frac{k_{Fp} (X)}{k_{Fp} (^{12}C)}-1\right)
+b_2\left(\frac{k_{Fn} (X)}{k_{Fn} (^{12}C)}-1\right)
+b_3\left(\frac{m_N^* (X)}{m_N^* (^{12}C)}-1\right),
\label{proposed_fit}
\end{equation}

The parametrization consists of three terms. The first term is
dependent on the proton content of the nucleus by the proton Fermi
momentum, the second term is proportional to the neutron Fermi
momentum, and the third term accounts for effective-mass effects
arising from the RMF framework. In a simple Fermi gas model, this last
contribution, associated with the coefficient $b_3$, would not be
required.
\begin{figure}[h!]
\centering
\includegraphics[scale=0.45]{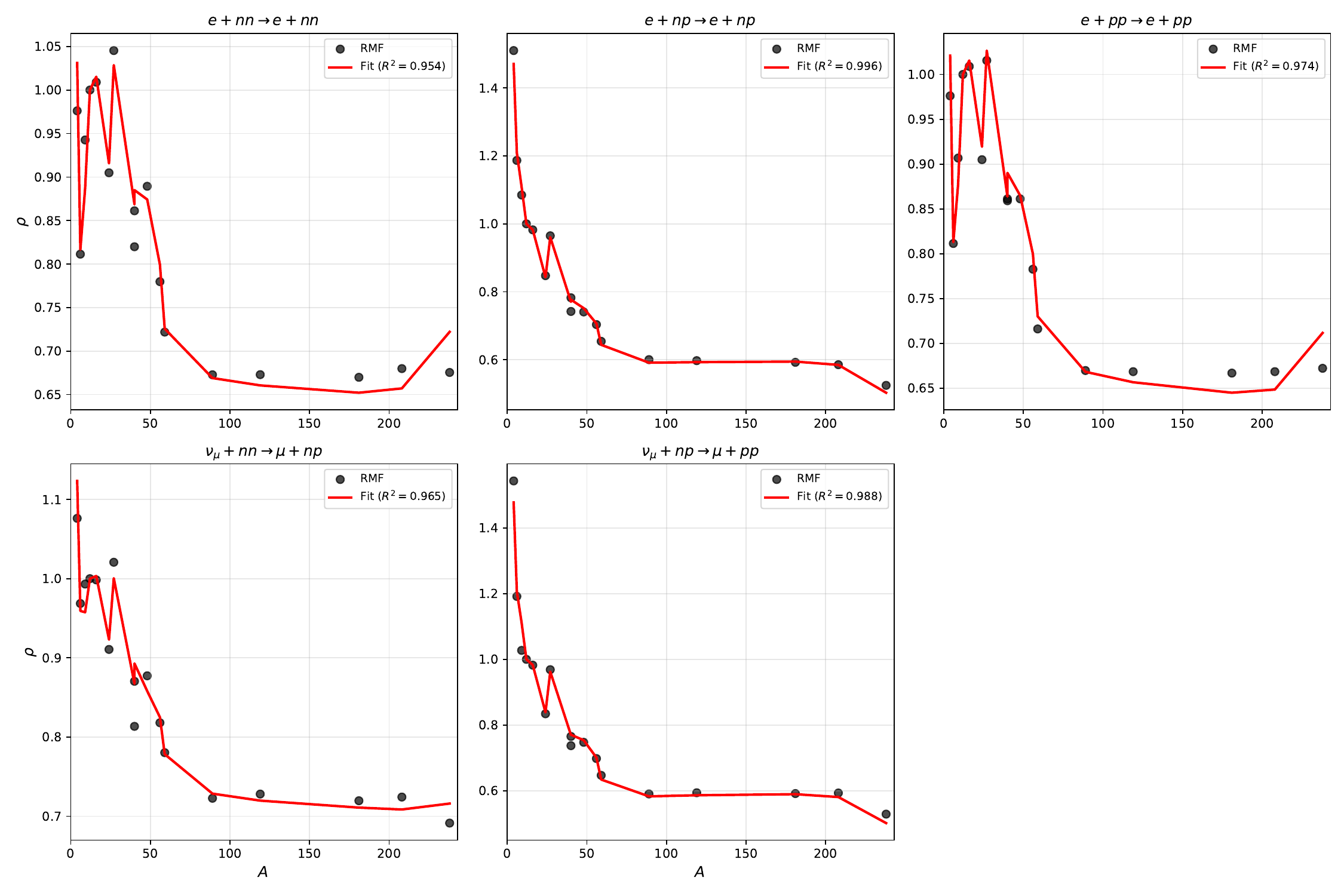}
\caption{Comparison between the factorized fit (solid lines) and the
microscopic RMF calculations as a function of the mass number
$A$ for the five emission channels considered. The fit is performed for
$\rho(X)$ after dividing out the volume and reduced phase-space factors,
see Eq.~\eqref{definiciones}.}
\label{figglobalfit}
\end{figure}

\begin{table*}[t]
\caption{Parameters of the reduced-response scaling function
$\rho(X)$ for the different emission channels. The last column reports
the coefficient of determination $R^2$ of the linear
regression.}
\label{fit_parameters_3p}
\begin{tabular*}{\textwidth}{l@{\extracolsep{\fill}}ccccc}
\toprule
Channel &
$b_1$ &
$b_2$ &
$b_3$ &
$R^2$ \\
\midrule
$e + nn \rightarrow e + nn$
 & $0.77$ & $-0.09$ & $1.84$ & $0.95$ \\

$e + np \rightarrow e + np$
 & $-0.58$ & $-0.20$ & $1.95$ & $0.99$ \\

$e + pp \rightarrow e + pp$
 & $0.90$ & $-0.20$ & $1.78$ & $0.97$ \\

$\nu_\mu + nn \rightarrow \mu + np$
 & $0.43$ & $-0.28$ & $1.34$ & $0.96$ \\

$\nu_\mu + np \rightarrow \mu + pp$
 & $-0.64$ & $-0.13$ & $2.03$ & $0.99$ \\
\bottomrule
\end{tabular*}
\end{table*}

We use $^{12}$C as the reference nucleus, fixing
$k_{Fp}(^{12}\mathrm{C}) = k_{Fn}(^{12}\mathrm{C}) =
225~\mathrm{MeV}/c$ and $m^*(^{12}\mathrm{C})/m_N = 0.8$. We also
remark that the effective $\Delta$ mass, $m^*_\Delta$, reported in
Table~\ref{parametros}, is an input of the microscopic RMF calculation
and is not treated as an independent scaling variable. Its nuclear
dependence is assumed to follow the same scaling behavior as the
effective nucleon mass.

The fitted coefficients reported in Table~\ref{fit_parameters_3p}
exhibit a clear dependence on the emission channel, reflecting the
different dynamical mechanisms that govern each reaction. The relative
weight of the $k_{Fp}$ and $k_{Fn}$ terms varies across channels,
effectively encoding the sensitivity to proton-neutron imbalance
through the different evolution of the two Fermi momenta with nuclear
number. The effective-mass coefficient $b_3$ contributes with the same
sign and with a similar magnitude in all cases, highlighting the role
of RMF medium effects.

\begin{figure}[h!]
\centering
\includegraphics[scale=0.45]{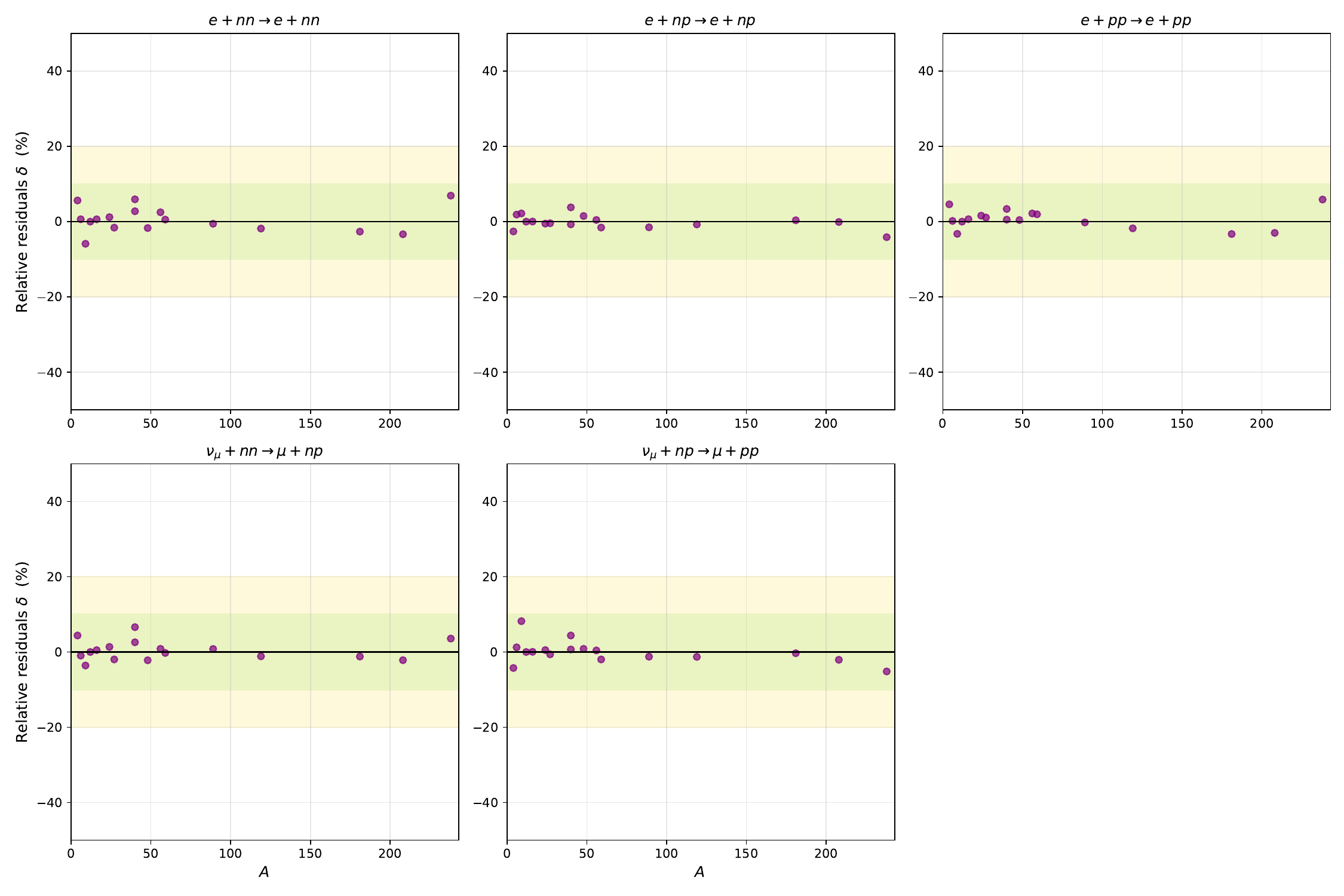}
\caption{Relative residuals as a function of the mass number $A$ for
the five emission channels~\mbox{considered}. The purple circles represent the different nuclei analyzed in this work. The green shaded area indicates a relative residual interval of $\pm 10\%$, while the yellow shaded area corresponds to a $\pm 20\%$ relative residual interval.}
\label{residuos}
\end{figure}

The values of $R^2$ (the coefficient of determination of the linear
regression) reported in the table provide a global measure of
the quality of the parametrization for each channel. The results demonstrate a uniformly high quality of the fit accoss all channels, with $R^2$ above 0.95. In Fig.~\ref{figglobalfit},
we display the results of the separated channel fit as a function of the mass
number $A$. We observe that, for all channels, the factor $\rho$ tends
to decrease with increasing nuclear size and approaches a nearly
constant value for heavy nuclei. The figure also illustrates the
relatively very good
description in all channel. The $pp$ channel with $R^2=0.99$, is particularly relevant
for neutrino-induced reactions, where it provides the dominant
contribution.

It is important to emphasize,
however, that the present parametrization is not intended to reproduce
the microscopic 2p2h response of a specific nucleus-which is obtained
from the underlying RMF calculations—but rather to describe how this
response evolves across different nuclear targets over a wide mass
range. To quantify the quality of the fit over this nuclear range, a
quantitative validation is therefore performed through the analysis of
the relative residuals shown in Fig.~\ref{residuos}, defined as

\begin{equation}
\delta(\%) =
\frac{Y^{\mathrm{RMF}} - Y^{\mathrm{fit}}}
     {Y^{\mathrm{RMF}}}
\times 100 .
\label{residuals}
\end{equation}

The resulting parametrization reproduces the main trends of the
scaling factors across different nuclei and emission channels. As
shown by the residual analysis, the relative residuals remain
typically within $\pm 10\%$, corresponding to relative deviations of
the order of $10\%$. As shown in Fig.~\ref{residuos}, the relative
residuals display the largest dispersion for the lightest nuclei,
where sizable positive and negative deviations are observed. For
intermediate-mass systems, the residuals cluster more tightly around
zero, with most values lying within the $5\%$ band and
exhibiting a smooth dependence on nuclear size. For the heaviest
nuclei, the dispersion increases again, although it remains less
pronounced than in the lightest systems and does not show a strong
systematic bias.

\section{Conclusions}
\label{conclusiones}

In this work we have investigated the nuclear dependence of
two-particle--two-hole meson-exchange current contributions to
inclusive electron--nucleus and neutrino-nucleus scattering within
the relativistic mean-field framework, extending microscopic
calculations from carbon to a broad set of nuclear targets ranging
from light to heavy systems.

We find that the dominant nuclear dependence of the transverse
two-particle responses can be largely absorbed into a
nucleus-dependent scaling factor relative to carbon. For most
medium-mass nuclei, the response shapes are similar in the vicinity of
the peak, while departures from perfect scaling are mainly observed
for the lightest and heaviest systems and at large energy transfers,
where subleading nuclear effects become more relevant.

On the basis of these observations, we adopt a factorized form for the
scaling ratio, separating volume and reduced phase-space contributions
and collecting the remaining nuclear dependence in a reduced-response
scaling function $\rho(X)$. We parametrize $\rho(X)$ using the proton
and neutron Fermi momenta and the RMF effective mass, obtaining a
compact representation of the microscopic results across nuclei rather
than a statistical fit in the strict sense.

The overall quality of the parametrization is moderate, capturing the
main trends of the scaling ratios across the nuclear chart. For
medium-mass nuclei, the relative deviations are typically bellow the $\sim
10\%$ level, as can be inferred from the residual analysis. Larger
deviations are concentrated in the lightest and heaviest systems,
reflecting both genuine nuclear-structure effects and the limitations
imposed by the restricted set of microscopic calculations considered
in this work, which spans 16 nuclei. These results illustrate that
achieving a fully universal description across all nuclei and emission
channels remains a challenging task within the present dataset.

Despite these limitations, the proposed scaling strategy provides a
practical and transparent framework to extrapolate microscopic RMF
calculations from a carbon reference to other nuclear targets of
experimental interest, including neutron-rich nuclei. Its simplicity
makes it well suited for exploratory implementations in neutrino event
generators, where it can support a more controlled treatment of
two-particle--two-hole contributions.

In future work, we plan to apply the same factorized scaling
prescription to antineutrino-induced reactions. We will also explore
comparisons with other microscopic 2p2h models, extend the set of
benchmark nuclei, and confront the results with available experimental
data. These steps will help to provide a more reliable and controlled
input for neutrino oscillation experiments, in particular for analyses
involving heavy targets.

\appendix
\section{Transverse electromagnetic response for the $nn$, $np$, and $pp$ channels}
\label{apendice1}

In this appendix we present the transverse electromagnetic responses
$R^T_{\rm em}$ and the corresponding scaling ratios for
electron-induced reactions. These results are obtained using the same
scaling procedure discussed in the main text, although they lead to
ratios that differ from the neutrino case due to the different isospin
structure of the electromagnetic and weak charged currents.

\begin{figure}[h]
\centering
\includegraphics[scale=0.95]{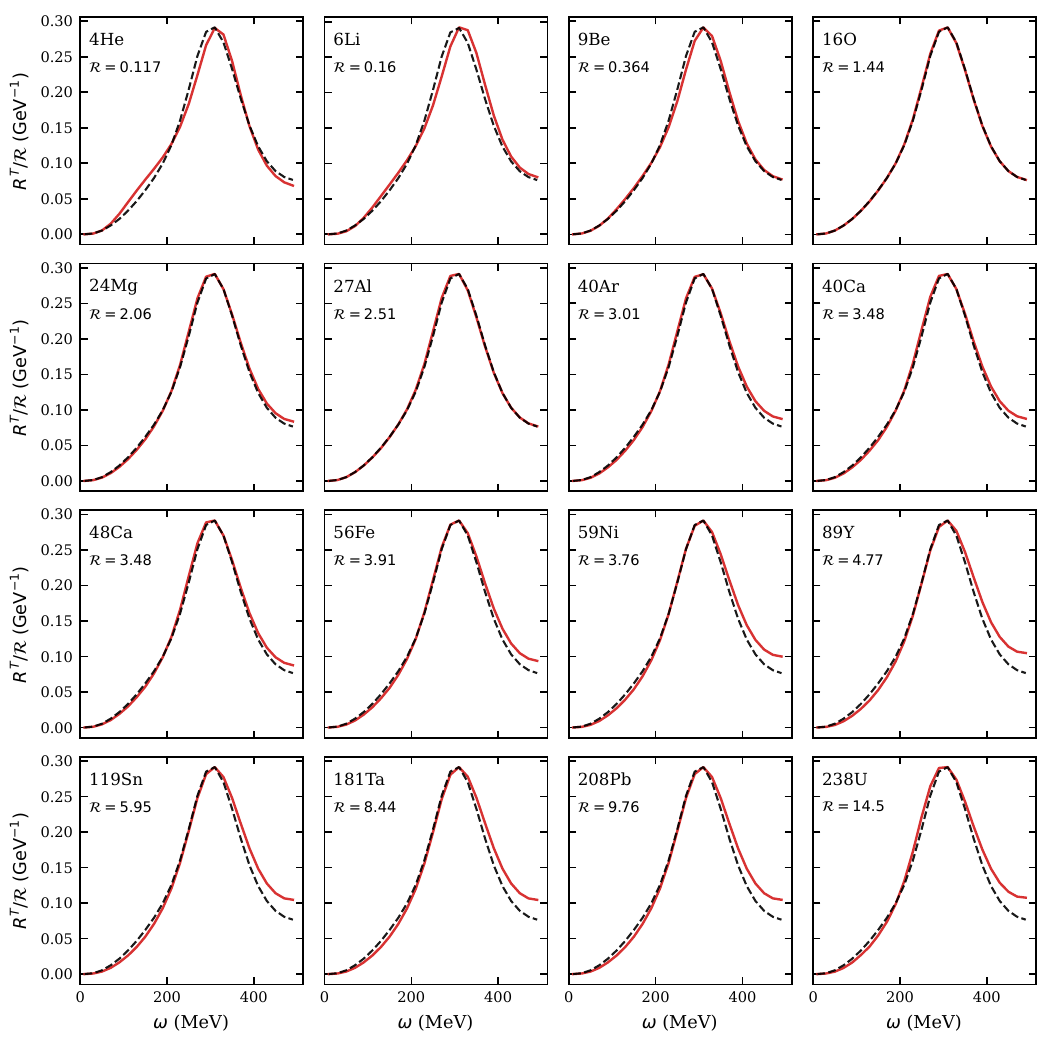}
\caption{Transverse 2p2h responses for electron reaction
$pp \rightarrow pp$ emission at $q = 500$~MeV/$c$.
Solid colored lines correspond to each nucleus, while the dashed black
line shows the ${}^{12}$C result. Each panel displays the response
rescaled by the factor $\mathcal{R}$ indicated in the
figure.}
\label{fig_e_pp}
\end{figure}

\begin{figure}[h]
\centering
\includegraphics[scale=0.95]{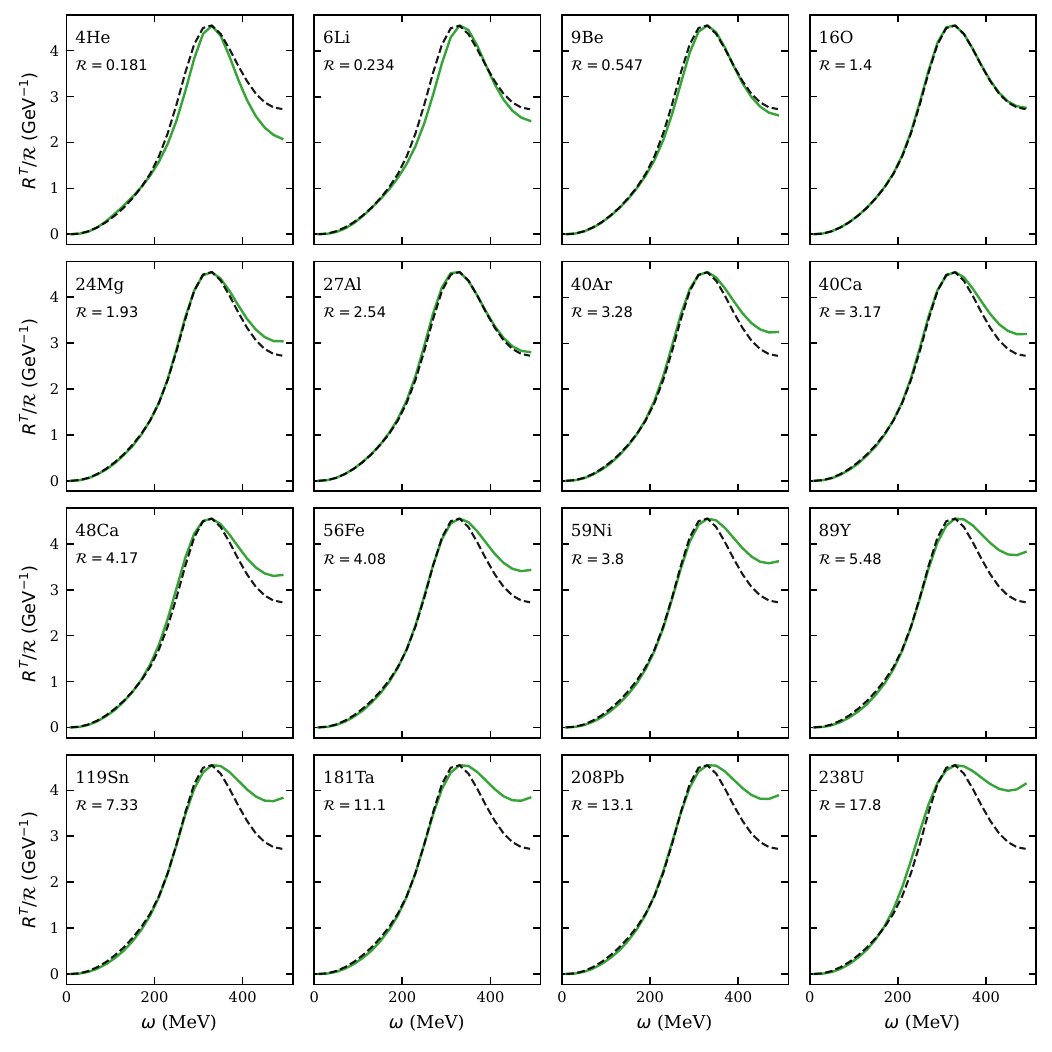}
\caption{Same as Fig.~\ref{fig_e_pp}, but for
$np \rightarrow np$ emission.}
\label{fig_e_np}
\end{figure}

\begin{figure}[h]
\centering
\includegraphics[scale=0.95]{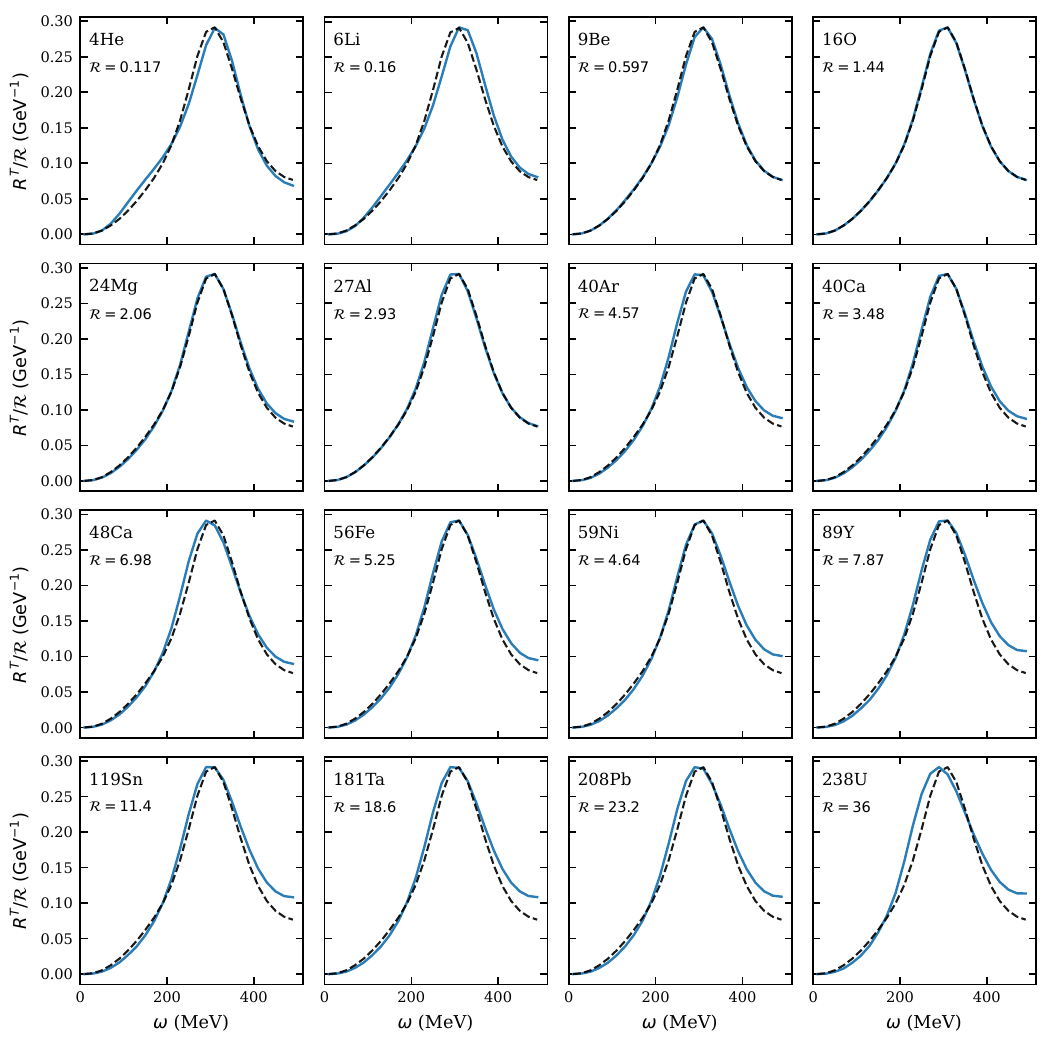}
\caption{Same as Fig.~\ref{fig_e_pp}, but for
$nn \rightarrow nn$ emission.}
\label{fig_e_nn}
\end{figure}

The analysis of the $nn$, $np$, and $pp$ channels in electron
scattering reveals systematic trends and allows for a direct
comparison with neutrino-induced reactions. In Figures~\ref{fig_e_pp},
\ref{fig_e_np}, and \ref{fig_e_nn} display the transverse responses
for the $pp$, $np$, and $nn$ emission channels, respectively. In all
cases, the same scaling prescription employed in the main analysis
provides a satisfactory description for medium-mass nuclei.  As in the
neutrino case, larger deviations are observed for the lightest and
heaviest systems, particularly at energy transfers above the peak.\\

It is worth noting that the scaling factors extracted for the
electron-induced $nn \rightarrow nn$ channel exhibit both a magnitude
and a growth with nuclear mass very similar to those found for the
neutrino-induced $nn \rightarrow np$ channel. A comparable
correspondence is also observed between the $np \rightarrow np$
channel in electron scattering and the $np \rightarrow pp$ channel in
neutrino scattering. This does not imply that the scaling ratio
$\mathcal{R}$ is solely determined by the isospin configuration of the
initial nucleon pair, but rather indicates that it also retains
sensitivity to the detailed dynamics of each emission channel.

 ------------------------------------------------------------
\begin{acknowledgments}
This work was supported by Grant No. PID2023-147072NB-I00 funded by
MICIU/AEI/10.13039/501100011033 and by the ERDF/EU, and by Grant
No. FQM-225 funded by the Junta de Andalucía. Additional financial
support was provided by the Spanish Ministry of Science, Innovation
and Universities under Grant No. PID2022-140440NB-C22, and by the
Junta de Andalucía under Contracts PAIDI FQM-370 and PCI+D+i
``Tecnologías avanzadas para la exploración del universo y sus
componentes'' (Code AST22-0001)..
\end{acknowledgments}
\begingroup
    \renewcommand{\section}[2]{}

\endgroup

\end{document}